\documentclass[12pt]{article}

\usepackage{amsmath,amssymb} 

\newcommand{\thet}{\theta}
\newcommand{\e}{\epsilon}
\newcommand{\ct}{a_j^\dagger(\thet)}
\newcommand{\at}{a_j(\thet)}
\newcommand{\alt}{\alpha_j(\thet)}
\newcommand{\clt}{\alpha_j^\dagger(\thet)}
\newcommand{\At}{{\cal A}_\thet}
\newcommand{\Ct}{\hat {\cal A}_\thet}
\newcommand{\Bphi}{{\cal B}_\phi}
\newcommand{\Cphi}{\hat {\cal B}_\phi}
\newcommand{\W}{W(M,N)}
\newcommand{\Ham}{{\cal H}}
\newcommand{\HIt}{H_I^r([\thet])}

\begin{document}

\title{\bf Quantum Mechanical Description of A Quasi-Static Process 
and An Application to Change of Temperatures
}

\author{Tsunehiro Kobayashi\footnote{E-mail: 
kobayash@a.tsukuba-tech.ac.jp}
 \\ 
{\footnotesize\it Department of General Education 
for the Hearing Impaired,}
{\footnotesize\it Tsukuba College of Technology}\\
{\footnotesize\it Ibaraki 305-0005, Japan}}

\maketitle

\begin{abstract}
A quasi-static process is realized in a purely quantum-mechanical 
model which is described by oscillator (or particle) systems having 
relative-phase interactions. 
Time development of a mixture of two oscillator (or particle) systems 
which have different temperatures is investigated in the quantum mechanical 
framework. 
We study how they go to a thermal equilibrium. 
Time dependence of the temperature of the object is numerically studied 
and the difference of the time dependence between classical objects and 
quantum objects is clearly pointed out. 

\end{abstract}

\thispagestyle{empty}

\setcounter{page}{0}

\pagebreak

\section{Introduction} \label{sect.1.0}
 
 In the papers published recently [1,2] we have proposed a quantum mechanical 
 description of thermal equilibrium states in terms of eigenstates of 
 relative-phase interactions $H^r_I([\theta])$ between oscillators (or 
 particles) with phases which are introduced as $a_j(\theta)=a_je^{-i\theta_j}$, 
 where $a_j$ is the annihilation operator of the $j$th oscillator 
 (or particle).  
 The interaction projects the physical space ${\cal H}_0$ of the original 
 Hamiltonian 
 $H_0=\e \sum_{j=1}^N  a_j^\dagger a_j$, 
 which is spanned by the direct products of the usual number states 
 $\prod_{j=1}^N|n_j>$, onto the thermal subspace ${\cal H}_{thermal}$ 
 of ${\cal H}_0$, where all the eigenstates having the same energy eigenvalue 
 have the same probability as if the fundamental principle of 
 thermal equilibrium, the principle of equal {\it a priori} probability, 
 is satisfied. 
 It has been shown that the thermal subspaces characterized by 
 the different phases are equivalent as the thermal subspace of 
 ${\cal H}_0$ [3]. 
 The microcanonical ensemble is derived by the average operation with respect 
 to the whole of the equivalent subspaces which cover the original physical 
 space $\Ham_0$ completely. 
 In the model it has been shown that the effective density-matrix 
 for arbitrary small subsystems of the total system are approximated 
 by canonical ensemble of usual statistical mechanics [4]. 
 In the procedure a common temperature is uniquely introduced 
 and determined by 
 the physical quantities of the heat-bath which is taken as the remaining 
 part of the total system except the small subsystems. 
 We can also see that this model can be described by using new operators 
 defined on the thermal subspace $\Ham_{thermal}$ in very simple forms. 
 Furthermore in this description we can easily see that the relative-phase 
 interaction provides transitions only on the thermal subspace. 
 This means that quasi-static processes [4] 
 can be defined as quantum processes. 
 In this paper we shall show the above results and apply it to the calculation 
 of change of temperatures. 
 
 In $\S$2 we briefly review the model and derive useful operators on the 
 thermal subspace. 
 The interactions of two systems and quasi-static processes are investigated 
 in $\S$3. 
 In $\S$4 time evolutions of the two systems are discussed. 
 Time evolutions of the temperatures of the interacting systems are 
 calculated in $\S$5. Remarks on the model are presented in $\S$6.

 \section{Basic ideas of the model and useful operators on $\Ham_{thermal}$} 
\label{sect.2.0}

In order to make our discussions clarify we review on the previous work 
describing thermal equilibrium for an oscillator system with one energy 
scale $\e$. 
(For details, see Ref. 1.) 
We shall also introduce useful operators on the thermal subspace. 

\hfil\break
{\it 2-1. Model} 
 
 The model is described by the system composed of $N$ oscillator (or $N$ 
 particle) which are described by the Hamiltonian 
 $\Ham_0=\e \sum_{j=1}^N a_j^\dagger a_j$, 
 where $a_j^\dagger$ and $a_j$ are, respectively, the creation and 
 annihilation operators and follow the commutation relations 
 $[a_j,a_j^\dagger]=\delta_{ij}$. 
 The phases are introduced in terms of the replacement of the operators 
 with the new ones 
 $a_j(\theta)=a_je^{-i\theta_j}$ and $a_j^\dagger(\theta)=a_j^\dagger e^{i\theta_j}$. 
 In such a replacement $H_0$ and the commutation relations do not change. 
 Using the usual number states satisfying 
 $ a_j^\dagger a_j|n_j>=n_j|n_j>$ and $a_j|0>=0$, we can obtain 
 the eigenstates of $H_0$ with the eigenvalues 
 $E_M=\e M$ $(M=0,1,2,...)$ as follows; 
 \begin{equation}
 |M;[n_j],[\theta_j]>\equiv |M;n_1,n_2,\cdots,n_N;\theta_1,\theta_2,\cdots,\theta_N>
       =\prod_{j=1}^N|n_j,\theta_j>\delta_{\sum_{k=1}^N n_k,M} \ ,
 \end{equation} 
 where $|n_j,\theta_j>=|n_j>e^{in_j\theta_j}$. 
 Note that we have the equation 
 $\ct |n_j,\theta_j>=\sqrt{n_j+1}|n_j+1,\theta_j>, \ 
 \at|n_j,\theta_j>=\sqrt{n_j}|n_j-1,\theta_j>$ and 
 $\ct\at|n_j,\theta_j>=n_j|n_j,\theta>$. 
 States in which all the eigenstates with the same energy eigenvalue 
 have the same probability, as if the principle of equal {\it a priori} 
 probability with respect to the eigenstates of $H_0$ is satisfied, 
 are given as 
 \begin{equation} 
 |M,N,[\theta]>=\sum_{p[n_j]} |M;[n_j],[\theta_j]>/\sqrt{W(M,N)},
 \end{equation}
 where the sum should be taken over all the different combinations of 
 $[n_j]=(n_1,n_2,...,n_N)$ and the number of the combinations 
 $W(M,N)$ is given by 
 $W(M,N) ={(M+N-1)! \over M!(N-1)!}$. 
 
 \hfil\break
 {\it 2-2. Relative-phase interaction} 
 
 We introduce the following operators [1-3];
\begin{equation}
\clt\equiv \ct(\sqrt{\hat N_j+1})^{-1},\ \ 
\alt \equiv (\sqrt{\hat N_j+1})^{-1} \at 
\ \ {\rm for} \ j=1,...,N,
\end{equation} 
where $\hat N_j=\ct \at$ is the excitation-number operator for the $j$th 
oscillator. 
We see that 
\begin{align}
\clt|n_j,\theta_j>&=|n_j+1,\theta_j>,\ \ {\rm for}\ \forall n_j=0,1,2,...&\nonumber \\
\alt|n_j,\theta_j>&=|n_j-1,\theta_j>,\ \ {\rm for}\ \forall n_j\geq1&
\end{align} 
and $\alt|0>_j=0$. 
The relative-phase interaction with the energy scale $\e_g$is written as 
\begin{equation} 
\HIt=\e_g{\hat N \over N}\sum_{j=0}^N({1 \over N} \sum_{k=0}^N\alpha_j^\dagger
\alpha_k e^{i\theta_{jk}}+|0>_j\ _j<0|),
\end{equation} 
where $\alpha_j^\dagger=\alpha_j^\dagger(0),\ \alpha_k=\alpha_k(0), 
\hat N=\sum_{j=1}^N\hat N_j$ (the total excitation-number operator) and 
the relative phase 
$\theta_{jk}=\theta_j-\theta_k$. 
We obtain the relation
\begin{equation} 
H|M,N,[\theta]>=(\e+\e_g)M|M,N,[\theta]>
\end{equation} 
for $H=H_0+\HIt$. 
We see that the physical space of the total Hamiltonian $H$ 
is spanned by the set of the states $|M,N,[\theta]>$ for $M=0,1,2,...$, which is nothing but the thermal subspace $\Ham_{thermal}$ of the original space $\Ham_0$. 
Thermal limit will be described by the limit $\e_g \rightarrow 0$, 
where the eigenvalues of $H$ can approximate to those of $H_0$. 
Note here that the states with the different phase parameters $|M,N,[\theta']>$ 
are not the eigenstates of $\HIt$, that is, 
$\HIt|M,N,[\theta']>\not=\e_gM|M,N,[\theta']>$ for $[\theta']\not=[\theta]$. 

\hfil\break
{\it 2-3. Derivation of canonical ensemble and temperature} 

We divide the total system into two groups, 
$b_j\equiv a_j$ ($j=1,2,...,N_b)$) ($b$-group) and 
$d_{j-N_b}\equiv a_j$ for $j\geq N_b+1$ ($d$-group). 
The eigenstates of the total Hamiltonian $|M,N,[\theta]>$ 
can be written in terms of the direct product of the eigenstates of the groups 
such that 
\begin{align}
|M,N,[\theta][\phi]>&=
\sum_{M_b=0}^M \sum_{M_d=0}^M\delta_{M_b+M_d,M}
{\sqrt{W(M_b,N_b)} \over \sqrt{\W}} |M_b,N_b,[\theta]>_b& \nonumber \\
&\otimes 
\sqrt{W(M_d,N_d)} |M_d,N_d,[\phi]>_d,&
\end{align} 
where $|M_B,N_b,[\theta]>_b$ and $|M_d,N_d,[\phi]>_d$ 
are, respectively, the eigenstates on the thermal subspace of the $b$-group 
and that of the $d$-group. 
Hereafter we shall treat the $b$-group as the heat bath and the $d$-group 
as the object. 
Then the $b$-group is taken to be much lager than the $d$-group, 
i.e., $N_b>>N_d$ and $M_b>>M_d$. 

In order to evaluate the density matrix for the object ($d$-group) 
we have to perform the partial trace operation (Tr$_I$), 
because any quantum numbers of the heat-bath ($b$-group) state are not measured 
in this processes. 
We obtain the effective density matrix for the object as follows; 
\begin{align} 
\rho_{eff}([\phi])&={\rm Tr}_I(|M,N,[\theta][\phi]><M,N,[\theta][\phi]|),&\nonumber \\
&=\sum_{M_b=0}^M \sum_{M_d=0}^M \delta_{M_b+M_d,M} 
{W(M_b,N_b)W(M_d,N_d) \over \W} 
|M_d,N_d,[\phi]>_d\ _d<M_d,N_d,[\phi]|.&
\end{align} 
In the limit where the interaction with the heat bath is small enough to ignore, i.e.,
$\e_g<<\e$, we may again introduce the average over the un-measurable phases of
 the object, $[\phi]$, because $H_0^{ob}=\e \sum_{k=1}^{N_1} d_k^\dagger d_k$ has 
no $[\phi]$-dependence. 
Thus we obtain the density matrix for the mixed states with respect to 
the eigenstates 
of $H_0^{ob}$ 
\begin{equation} 
\rho_{eff} = \sum_{M_b=0}^M \sum_{M_d=0}^M \delta_{M_b+M_d,M} 
{W(M_b,N_b) \over W(M,N)} 
\sum_{p[m_j]} |M_d;[m_j]>_d \ _d<M_d;[m_j]|.
\end {equation} 
Note here that in the above procedure the average operation with respect to 
the phases of the heat bath, $[\theta]$, 
are not carried out, while it must be done to 
derive microcanonocal ensemble for the heat bath as was discussed in Refs. 1-3. 
The canonical ensemble is realized at the maximum of the probability 
$P_M(M_b)=W(M_b,N_b)W(M-M_b,N-N_b)/\W$ 
for finding the state $|M_b,N_b,[\theta]>_b\otimes |M_d,N_d,[\phi]>_d$ 
in $|M,N,[\theta][\phi]>$ [4]. 
At the maximum we can estimate the change of the coefficients in the probability 
with respect to the change of the object energy, that is, 
$\Delta E^{ob}=\e \Delta M_d$ with $\Delta M_d\sim M_d<<M,\ M_b$, as 
\begin{equation} 
{W(M_b-\Delta M_d,N_b) \over W(M_b,N_b)} \sim ({M_b \over M=b + N_b })^{\Delta M_d}
=e^{-{1 \over \e} \Delta E^{ob} {\rm ln}(1+N_b/M_b)}.
\end{equation} 
Thus we can determine the temperature $T$ as 
\begin{equation} 
kT=\e ({\rm ln}(1+N_b/M_b))^{-1}.
\end{equation} 
It is obvious that the temperature is written down by using the quantities of 
the heat bath, that is, the mean excitation number $(M_b/N_b)$ of the heat bath. 
We obtain the density matrix for canonical ensemble in the limit 
$M\sim M_b,\ N\sim N_b \rightarrow \infty$ 
\begin{equation} 
\rho_{eff} \simeq \sum_{M_d=0}^\infty e^{-\beta E_{M_d}^{ob}} \sum_{p[m_j]} 
|M_d;[m_j]>_d \ _d<M_d;[m_j]|/Z_{N_d}.
\end{equation} 
As  for the details of the discussions presented in this section, 
please see $\S$5 of Ref. 1. 

\hfil\break
{\it 2-4. Useful operators on the thermal subspace} 

Here we shall present a pair of useful operators in the following discussions, 
which are defined by 
\begin{equation} 
\At \equiv \sqrt{\hat N+N} \sum_{j=1}^N \alt/N,\ \ \Ct \equiv \sum_{j=1}^N \bar \alpha_j(\theta)(
\sqrt{\hat N+ N})^{-1}, 
\end{equation} 
where $\bar \alpha_j(\theta) \equiv \ct\sqrt{\hat N_j+1}$. 
It should be noted that they satisfy the commutation relation $[\At,\Ct]=1$. 
We easily se that on the thermal subspace $\Ham_{thermal}$ 
\begin{align} 
\At|M,N,[\theta]>&=\sqrt{M} |M-1,N,[\theta]>,& \nonumber \\ 
\Ct|M,N,[\theta]>&=\sqrt{M+1}|M+1,N,[\theta]>&
\end{align} 
and $\Ct\At|M,N,[\theta]>=M|M,N,[\theta]>$. 
We may say that they each other behave as hermitian conjugate operators 
on $\Ham_{thermal}$. 
Now we can write the relative-phase interaction and the eigenstates 
in very simple forms on $\Ham_{thermal}$ as follows; 
\begin{equation} 
\HIt=\e_g\Ct\At, \ \ |M,N,[\theta]>={1 \over \sqrt{M!}} (\Ct)^M|0>.
\end{equation} 
An important feature of the new operators can be seen in the choice of $N=1$, 
where they coincide with the original operators such that 
$\At=a(\theta)$ and $\Ct=a^\dagger(\theta)$. 

\section{Interaction between two systems and quasi-static process} 
\label{sect.3.0}

Let us consider interactions between two oscillator systems composed of 
the same oscillators (or particles). 
In order to distinguish them one is called $a$-system described by 
the original Hamiltonian 
$H_0^a=\e\sum_{j=1}^{N_a} a_j^\dagger a_j$ and the other $b$-system described by 
$H_0^b=\e\sum_{j=1}^{N_b} b_j^\dagger b_j$. 
Hereafter  we shall represent the operators defined in {\it 2-4} by $\At$ 
for the $a$-system and by $\Bphi$ for the $b$-system. 
We postulate that the $a$-system is in the thermal equilibrium on 
$\Ham_{thermal}^a$ and the $b$-system in that on $\Ham_{thermal}^b$ 
before the two systems are mixed. 
That is to say, the $a$- and $b$-systems are, respectively, described by the 
states $|M_a,N_a,[\theta]>={1 \over \sqrt{M_a!}} (\Ct)^{M_a}|0>_a$ and 
$|M_b,N_b,[\phi]>={1 \over \sqrt{M_b!}} (\Cphi)^{M_b}|0>_b$. 
The thermal interaction may be written as follows; 
\begin{equation} 
H_I=\e_g \hat {\cal A}_T {\cal A}_T,
\end{equation} 
Where $\hat {\cal A}_T=\Ct +\Cphi,\ {\cal A}_T=\At +\Bphi$. 
The terms $H_I^a([\theta])=\e_g\Ct\At$ and 
$H_I^b([\phi])=\e_g\Cphi\Bphi$ 
involved in $H_I$ are, respectively, nothing but 
the relative-phase interactions 
for the $a$- and $b$- systems, 
which are already taken into account in the description of the states 
$|M_a,N_a,[\theta]>$ and $|M_b,N_b,[\phi]>$. 
Then the interaction between the $a$- abd $b$- systems is represented by 
\begin{equation} 
H_I^{ab}=\e_g(\Ct\Bphi + \Cphi\At). 
\end{equation} 
From the relations of (14) we see that this interaction induces 
the transitions 
only among the states on the direct-product space of two thermal subspaces 
for the two systems such as 
$|M_a,N_a,[\theta]>\otimes |M_b,N_b,[\phi]>$ 
on $\Ham_{thermal}^a\otimes \Ham_{thermal}^b$, and then  it preserves both of 
the thermal equilibrium for the $a$- and $b$-systems, 
even though the thermal equilibrium of the total system is not realized. 
It means that the processes induced by this interaction represent 
quasi-static processes. 
Then we may say that all thermodynamical quantities of the two systems, 
such as the temperatures of $a$- and $b$-systems, chemical potentials and so on, 
are well-defined physical quantities at arbitrary moments of 
this physically changing process. 

\section{Time evolutions of the two systems} \label{sect.4.0}

In the interaction picture the unitary time evolution operator is evaluated by the formula 
\begin{equation}
U(t,0)={\rm exp}[-i\int_0^tH_{int}(t')dt'/\hbar]
\end{equation}
where 
\begin{align} 
           H_{int}&= e^{i(H^a+H^b)t} H_I^{ab} e^{-i(H^a+H^b)t} & 
           \nonumber \\
&= \e_g(\Ct(t) \Bphi(t) +\Cphi(t) \At(t)),&
\end{align} 
where $H^a=H_0^a+H_I^a([\thet]),\ \At(t)=\At  e^{-i(\e+\e_g)t/\hbar}, \Ct(t)=\Ct  e^{i(\e+\e_g)t /\hbar}$ and so on. 
Note  here that the $t$-dependence drops from $H_{int}(t)$ because the operators $\At(t)$ and 
$\Bphi(t)$  have the same time dependence and then $U(t,0)$ is the exact formula because 
$[H_{int}(t'),H_{int}(t'')]=0$ is satisfied for arbitrary $t'$ and $t''$. 
Putting $\omega_g=\e_g/\hbar$, we obtain 
\begin{equation}
U(t,0)={\rm exp}[-i\omega_g(\Ct \Bphi+\Cphi \Ct)t]. 
\end{equation} 
We study the case where two systems are mixed at $t=0$ and the initial state is given by 
$$ 
|\Psi_I>=|M_a,N_a,[\thet]>\otimes |M_b,N_b,[\phi]>. 
$$ 
We should study two different situations; one is expressed by the interaction between 
two systems which have a comparable size, that is, $N_a\simeq N_b$, 
and the other is the case where one of the two (heat bath) is extraordinary lager than 
the other (object), e.g., $N_a>>N_b,\ M_a>>M_b$. 
 
{\it Case (1) for $N_a \sim N_b$}: 
We may consider that all the constituents in the both systems couple each other. 
The density matrix is written down as 
$\rho=U(t,0)|\Psi_I>\ <\Psi_I|U(t,0)^\dagger$ 
and the effective density matrix for the object($b$-system) is obtained by performing the 
partial trace operation (Tr$_I$) for the variables of the other ($a$-system) as 
\begin{equation} 
\rho_{eff}=\sum_{m=0}^M|c_m(t)|^2|m,N_b,[\phi]>\ <m,N_b,[\phi]|, 
\end{equation} 
where the coefficients $c_m(t)$ are defined by 
$U(t,0)|\Psi_I>=\sum_{m=0}^M c_m(t)|M-m,N_a,[\thet]>\otimes |m,N_b,[\phi]>$ 
with the constraints $\sum_{m=0}^\infty |c_m(t)|^2=1$ and 
$M=M_a+M_b$. 
The effective density matrix $\rho_{eff}$ is represented by a mixed state on $\Ham_{thermal}^b$. 
We should pay attention to the fact that the partial trace operation with respect to the variables of
 the non-objective matters ($a$-system) plays an essential role in deriving the mixed state property 
of $\rho_{eff}$. 
It is also note that $\rho_{eff}$ is not yet represented by any mixed states on $\Ham_0^b$.  
The decoherence with respect to the eigenstates of $H_0^b$, that is, the decoherence on $\Ham_0^b$, 
is accomplished by the average over the phases $[\phi]$ as noted in {\it 2-3}. 
In the following discussions, 
however, we have to use $\rho_{eff}$ because we cannot take the thermal limit $\e_g \rightarrow 0$, 
where any time evolution does not take place. 
 
We can study the time development of the total system 
in perturbation because of $\e>>\e_g$. 
The matrix elements at $t=\tau$ are obtained as follows; 
\begin{align} 
  <\Psi_I|U(\tau,0)|\Psi_I>&\simeq 1-{1 \over 2} (\omega_g \tau)^2(2M_aM_b+M_a+M_b),& \nonumber \\ 
<M_a+1|\ <M_b-1|U(\tau,0)|\Psi_I>&\simeq-i\omega_g\tau \sqrt{(M_a+1)M_b},& \nonumber \\
<M_a-1|\ <M_b+1|U(\tau,0)|\Psi_I>&\simeq -i\omega_g \tau \sqrt{(M_b+1)M_a}&
\end{align} 
and others are less than the order of $(\omega_g \tau)^2$, 
where $|M_a> |M_b>\equiv |M_a,N_a,[\thet]>\otimes |M_b,N_b,[\phi]>$ 
is used. 
The probabilities up to the order $O((\omega_g \tau)^2)$ are 
given by 
\begin{align}
      P(M_a,M_b)&\simeq 1-(\omega \tau)^2(2M_aM_b+M_a+M_b),& \nonumber \\
P(M_a+1,M_b-1)&\simeq (\omega_g\tau)^2(M_a+1)M_b,& \nonumber \\
P(M_a-1,M_b+1)&\simeq (\omega_g\tau)^2(M_b+1)M_a.& 
\end{align} 
We obtain the time dependence of the mean excitation number of the object as 
\begin{equation} 
<\Cphi\Bphi>\equiv M_b(1)\simeq M_b  -(\omega_g\tau)^2(M_b-M_a).
\end{equation} 
This result is quite reasonable because the energy of the object increases if it is smaller than 
that of the environment and {\it vice versa}. 
Furthermore the change of the energy stops when both of the energies become same. 
Let us estimate the time when the time evolution stops. 
The time evolution equation for the excitation numbers is obtained from the relations for the $n$th 
steps of the perturbation 
\begin{align} 
M_b(n)&=M_b(n-1)-(\omega_g\tau)^2[M_b(n-1)-M_a(n-1)],& \nonumber \\
M_a(n)&=M_a(n-1)-(\omega_g\tau)^2[M_a(n-1)-M_b(n-1)].& 
\end{align} 
We have the equations for  the difference $m(n)=M_b(n)-M_a(n)$ as follows; 
\begin{equation} 
\Delta m(n-1)=-2(\omega_g\tau)^2 m(n-1) \Delta n, 
\end{equation}
where $\Delta m(n-1)=m(n)-m(n-1)$ and $\Delta n=1$.  
Since we may treat $n$ as a continuous variable in the region $n>>1$, 
we derive the relation 
\begin{equation} 
m(n)\simeq (M_b-M_a) e^{-2\omega_g^2 \tau t_n}, 
\end{equation} 
where the time corresponding to the $n$th step $t_n\equiv \tau n$. 
From the above result and the relation for the energy conservation 
$M_a(n)+M_b(n)=M_a+M_b$ for $\forall n$, we get 
\begin{equation} 
M_b(t_n)={M_a+M_b \over 2} +{M_b-M_a \over 2} e^{-2\omega_g^2 \tau t_n}. 
\end{equation} 
The time when the thermal equilibrium is almost realized is that 
when the difference $m(n)/M_b$ approximates to $0$. 
Then we can estimate it from the relation $m(n) \sim 1<<M_a, \ M_b$ as 
\begin{equation} 
t_{thermal}^{(1)}\simeq {1 \over 2\omega_g^2\tau} {\rm ln}(|M_b-M_a|). 
\end{equation} 
Note that the exponential damping property of $M_b(t_n)$ for $t_n$ 
is same as that of classical damped oscillation. 
We may put $\omega_g\tau <(M_a+M_b)^{-1/2} $ because the consistency of the  
perturbation, that is, all the probabilities in (23) must be positive. 
We have 
\begin{equation} 
t_{thermal}^{(1)}>{\sqrt{M_a+M_b} \over 2\omega_g} {\rm ln}(|M_b-M_a|)
\simeq {\sqrt{(E_a+E_b)/\e} \over 2\omega_g} {\rm ln}(|E_b-E_a|/\e).
\end{equation} 
From (11) the temperature of the object after the $n$-perturbation steps is derived as 
\begin{equation} 
T_b(n)\simeq (\e/k)[{\rm ln}(1+2N_b\e/(E_a+E_b-(E_b-E_a)e^{-2\omega_g^2\tau t_n}))]^{-1}. 
\end{equation} 
Since the time ($t_{meas}$) required for measuring the energy with the accuracy $\Delta E\sim \e$ 
is much shorter than the perturbation time ($t_{pert}$), i.e., 
$t_{meas}\sim \hbar/\e<< t_{pert} \sim \hbar/\e_g$ 
because of $\e>>\e_g$, we have time enough to measure the temperature which is 
determined by the knowledge on $\Ham_{thermal}^b$ involved in $\Ham_0^b$. 
 
{\it Case (2) where an object is put in a huge heat bath}: 
We have to take account of the fact that the object interacts with a very small part of  
the heat bath. 
Therefore the heat bath must be divided into two parts and the part ($a$-system) 
which interacts with the object ($b$-system) is comparable to the size of the object, 
that is, $N_a\simeq N_b$, and very much smaller than the non-interacting part. 
Such a separation has already been carried out in {\it 2-3} and written 
in the canonical ensemble given in (12). 
The effective density matrix for the object can be obtained as follows; 
\begin{equation} 
\rho_{eff}=\sum_{M_a=0}^\infty \sum_{m=0}^{M_a+M_b} e^{-\beta(M_a+M_b-m)\e}
|c_m(t)|^2|m,N_b,[\phi]>\ <m,N_b<[\phi]|,
\end{equation} 
where $\beta=1/kT$ ($T$ is the temperature of the heat bath).  
Following the same perturbative treatment, 
we derive the change of the mean excitation number  
of  the object as 
\begin{equation} 
<\Cphi \Bphi>\simeq  M_b-(\omega_g \tau)^2(M_b-<M_a>),
\end{equation} 
where the expectation value for the excitation of the interacting  part 
is  evaluated as 
\begin{equation} 
<M_a>=\sum_{M_a=0}^\infty M_a e^{-\beta M_a \e} ={e^{\beta\e} \over 
(e^{\beta \e}-1)^2}. 
\end{equation} 
The difference of the result of (33) from that of (24) appears in the fact that 
$<M_a>$ in (33) does not change during the interaction with the object, 
while $M_a$ in (24) does.  
The change of the excitation number after the $n$-perturbative steps is obtained as follows; 
\begin{equation} 
M_b(n)=<M_a>+(M_b-<M_a>)((1-(1-(\omega_g\tau)^2)^n). 
\end{equation} 
In the limit of $n\rightarrow \infty$ we see that 
\begin{equation}  
{\rm lim}_{t_n\rightarrow  \infty} M_b(n)=<M_a>.
\end{equation} 
It means that the object goes to the thermal equilibrium provided by the heat bath. 
Since we may represent $M_b(n)$ by 
$M_b(t_n)\simeq <M_a>+(M_b-<M_a>)e^{-\omega_g^2\tau t_n} $ 
for large $n$, the time  when the thermal equilibrium is almost realized 
is estimated 
as 
\begin{equation} 
t_{thermal}^{(2)} \simeq {1 \over \omega_g^2\tau} {\rm ln}(|M_b-<M_a>|), 
\end{equation} 
which is just twice of the $t_{thermal}^{(1)}$ for case (1). 
This is quite reasonable because in this case the $a$-system is a part of 
the heat bath does not 
change the thermal properties, whereas the $a$-system changes the states 
towards the thermal equilibrium in case (1). 
That is to say, the change of the temperature of the $b$-system in this case 
must be twice as large as that of case (1). 
We shall easily be able to observe this difference in experiments. 
 
Note here that $M_b(t_n)$ damps exponentially with the time for 
$t_n>>1/\omega_g^2\tau$ like the classical damped oscillation, 
while it damps as Gaussian for $t_n<<1/\omega_g$. 
The perturbation approach used in this section is one of 
the weak coupling limits [5].

 \section{Some numerical considerations} \label{sect.5.0}

Here let us numerically study the time-dependence of the temperature 
which may be 
the most probable observation in experiments.  
We investigate a simple example for case (1), 
that is, we study the change of the temperature for the mixture of two same particle 
systems (gases) which have a similar weight ($N_a\simeq N_b$) but 
different temperatures ($M_a\not= M_b$) and bottled in the same-sized cubes 
(the volume=$l^3(cm^3)$). 
Since the energy spectrum of a constituent particle are derived as 
$E_{n_x,n_y,n_z}={(\pi\hbar)^2 \over 2ml^2} (n_x^2+n_y^2+n_z^2)$, 
where $m$ is the mass of the particle and $n_x,\ n_y, \ n_z$ 
are positive integers, all the energy levels are represented by the lowest energy ($E_0$) 
$\times$ integers and then we may take $E_0$ as the fundamental level spacing 
$\e$ used in the previous discussions. 
(Exactly speaking, we have to use the method for the model involving different energy scales 
presented in Ref. 2.) 
The lowest energy is obtained as  
\begin{equation} 
\e=E_0\simeq 0.25\times 10^{-30} {1 \over l^2} ({\rm erg}),
\end{equation} 
where the gas is taken as the hydrogen gas. 
The temperature is written down as 
\begin{equation} 
T(t)\simeq 0.2\times 10^{-14}{1 \over l^2} {1 \over {\rm ln} (1+N_b/M_b(t))}\ \ (^oK), 
\end{equation} 
where $M_b(t)$ is given by (28) and $t=\tau n$ is used. 
 
Let us study a case corresponding to our daily phenomena, say $l=1$cm and $T\sim 10^2\ (^oK)$. 
Since the factor $0.2\times 10^{-14}$ is very small, the approximation 
ln$(1+N_b/M_b(t))\simeq N_b/M_b(t)<<1$ must be realized even at very low temperatures. 
Thus we get 
\begin{equation} 
T(t)\simeq {1 \over 2} (T_a+T_b) +{1 \over 2} (T_b-T_a) e^{-2(\omega_g^2\tau)t}
\end{equation} 
for $t>>1/\omega_g^2\tau$,  where $T_a$ and $T_b$ are, respectively, the initial 
temperatures of the $a$- and $b$-systems. 
This formula is quite reasonable from the classical point of view. 
We can determine the parameter $\omega_g^2\tau$ from experimental data. 
 
Let us study the case for the quantum size, for instance, $l=10^{-8}$cm. 
We get 
\begin{equation} 
T_b(t)\simeq 20\times{1 \over {\rm ln}(1+N_b/M_b(t))}\ (^oK). 
\end{equation} 
The difference from the above result (40) is found only below the critical 
temperature 
$T_c\simeq 20/{\rm ln}2\ (\simeq29) (^oK)$, 
where the relation $N_b/M_b(t)>>1$ must be satisfied. 
We have 
\begin{equation}
T_b(t)\simeq T_b(f)+{1 \over 20} \times T_b(f)^2
{M_b-M_a \over M_a+M_b} e^{-2(\omega_g^2\tau)t} 
\end{equation} 
for $t>>1/\omega_g^2\tau$, where 
$T_b(f)\equiv \e/[k\ {\rm ln}((N_a+N_b)/(M_a+M_b))]$  
representing the temperature in the final thermal equilibrium may approximate 
to $20/{\rm ln}(2N_b/(M_a+M_b))$ because of $N_a\simeq N_b$. 
We see that the time-dependence is same as that of the classical case but 
the factor of the time-dependent term is quite different from that of the classical 
case. 
Note that the numeral $1/20$ in (42) has the size ($l$(cm))dependence such as 
$5\times10^{14} l^2$. 
The difference of the quantum case from the classical case can experimentally 
be seen by measuring the size dependence. 
For case (2) we can proceed the same argument by replacing $M_a$ with $<M_a>$. 
 
\section{Remarks} \label{sect.6.0}
 
We have showed that a simple non-thermal equilibrium phenomena can be 
treated in the present quantum mechanical framework for thermodynamics. 
It should be stressed that in this scheme
 we can proceed all evaluations in preserving quasi-static property. 
We can, therefore, determine temperatures of all subsystems at arbitrary time. 
Davies presented quantum theory of open systems [5]. 
In his approach, however, we have no definite way to determine the temperature 
of the object at arbitrary time. 
It is also note that in his approach the canonical distribution for the heat bath 
is put by hand, whereas the distribution is derived in the large heat-bath limit 
($M$ and $N\rightarrow \infty$) in the present scheme as was shown in (12). 
Exactly speaking, we can study the deviation from the canonical distribution for 
systems having not large $M$ and $N$. 
It is interesting that our average operation over the phases has the same role as 
that of the projection operator $P_0$ introduced in section 10.2 of Ref. 5, 
which projects the space of the closed system onto the subspace described 
by real diagonal matrices.
 Thus we see that the phases in our model play an essential role for the derivation 
of the microcanonical and canonical ensembles. 
(See $\S$2-3.) 
There has been no model in which temperatures of interacing systems are 
{\it quantum mechanically } calculable. 
In fact we need not to introduce any temperatures {\it a priori} in our scheme. 
This point is quite different from other approach, for instance, 
the model for an oscillator interacting with environment by Unruh and Zurek [6]. 
 
The magnitude of the relative-phase interaction is still unknown, 
One possible observation of the interaction 
in cold-neutron experiments has been proposed [3]. 
We can also estimate the strength of the relative-phase interaction by measuring  
$\omega_g^2\tau$ through the time development of temperatures. 
Whether its magnitude is universal or not will be a good test for 
the universality of the relative-phase interaction. 
We should, however, not forget about the fact that in the present scheme the viscosity
 introduced from interactions on surfaces 
(boundary of heat bath and object), 
which will disturb the thermal equilibrium of the object, is not taken into account. 
Those problems must be studied in more general 
formalism [5]. 

 \hfil\break
\large{\bf  References}
 
 \hfil\break
[1] T. Kobayashi, Phys. Letters {\bf  A207} (1995) 320 and {\bf A208} (1995) 381. \hfil\break
[2] T. Kobayahsi, Phys. Letters {\bf A210} (1996) 241. \hfil\break
[3] T. Kobayahsi, Phys. Letters {\bf A222} (1996) 26. \hfil\break
[4] See test books of statistical mechanics. \hfil\break
[5] E. B. Davies, {\it Quantum Theory of Open Systems}, Academic Press, London, 1976. \hfil\break
[6] W. G. Unruh and W. H. Zurek, Phys. Rev. {\bf D40} (1989) 1071. 
 
\end{document}